# A facile top-down synthesis of phase pure CuBi$_2$O$_4$ photocathode materials by low-energy sequential ball milling

**Syed Farid Uddin Farhad**[a,b*], **Mosammat Irin Naher**[a], **Nazmul Islam Tanvir**[a,b], **M. S. Quddus**[b,c], **S. A. Jahan**[c], Suravi Islam[a], Md Aftab Ali Shaikh[d,e]

[a]Energy Conversion and Storage Research Section, Industrial Physics Division, BCSIR Dhaka Laboratories, Dhaka 1205, BCSIR
[b]Central Analytical and Research Facilities (CARF), Dhaka 1205, BCSIR
[c]Institute of Glass and Ceramic Research and Testing (IGCRT), Dhaka 1205, BCSIR
[d]Bangladesh Council of Scientific and Industrial Research (BCSIR), Dhaka 1205, Bangladesh
[e]Department of Chemistry, University of Dhaka, Dhaka 1000, Bangladesh

[*]Corresponding author: s.f.u.farhad@bcsir.gov.bd; sf1878@my.bristol.ac.uk

**Graphical Abstract**

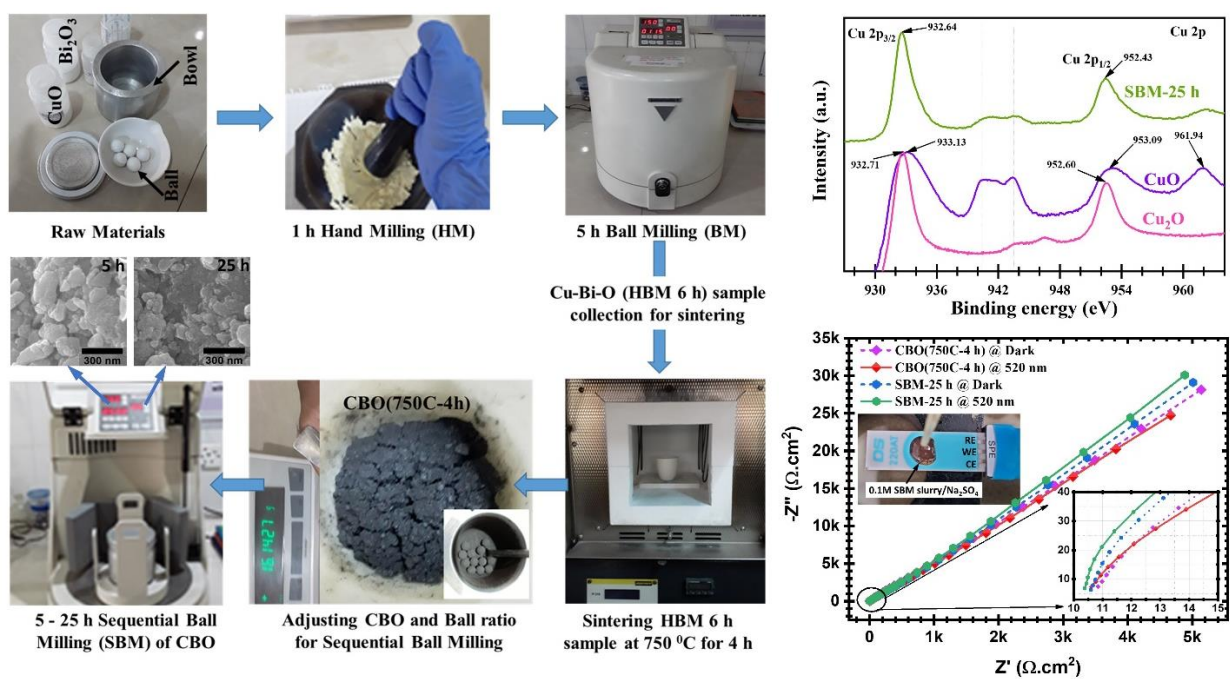

**Abstract**

Kusachiite (copper bismuth oxide) is one of the most promising photocathode materials for high solar-to-hydrogen production efficiency. Here we attempt to synthesize phase pure copper bismuth oxide (CuBi$_2$O$_4$) nanopowders using a facile solid-state reaction technique, subsequently sintered at ~750 $^0$C for 4 h in air. These CuBi$_2$O$_4$ (CBO) powders have been further sequentially ball-milled (SBM) up to 25 h to elucidate the milling duration effect on the optical bandgap of the ball-milled CuBi$_2$O$_4$ (SBM-CBO). The X-ray diffractometry (XRD), Raman, X-ray Photoelectron spectroscopy (XPS), and UV-VIS-NIR Diffuse reflection studies suggest that phase pure tetragonal CBO could be grown from raw CuO and Bi$_2$O$_3$ powders. The variations in morphology and chemical composition of CBO with increasing milling hours were examined using field emission scanning electron microscopy (FE-SEM) and Energy Dispersive X-ray (EDX) microanalysis, respectively. The optical bandgap of SBM-CBO powders was measured in the range of ~1.70–1.85 eV. Photoelectrochemical studies of SBM-CBO slurry atop the screen-printed electrode with aqueous $Na_2SO_4$ electrolyte suggest its suitability as photocathode materials for sunlight-driven hydrogen production.

**Research Highlights**

1. Facile synthesis of CuBi2O4 nanopowder by low-energy sequential ball milling.
2. Variation of milling duration yielded tetragonal CuBi2O4 with a tunable bandgap of ~1.70 –1.85 eV.
3. Phase purity of CuBi2O4 confirmed by combined analyses of XRD, Raman, and XPS.
4. Photoelectrochemical performance of CuBi2O4 slurry suggests its suitability as a photocathode.

## Introduction

Since the first industrial revolution, the development of civilization has been immensely dependent on fossil fuels (coal, oil, natural gas, etc.) as energy sources. These nonrenewable energy sources are not only limited and exhausting but also cause perennial damage to our environment by producing many harmful byproducts, such as $CO_2$. Therefore, finding renewable, eco-friendly, affordable, and long-term storable energy sources as an alternative to fossil fuels to satisfy ever-growing energy demand has become a pressing need in the era of the 4th industrial revolution (4IR). To this end, one of the most promising approaches could be utilizing the immense and ubiquitous energy reservoir - the sun and earth-abundant water to establish a sustainable solar energy conversion and storage system. Photoelectrochemical (PEC) water-splitting is such a system to overcome the worldwide energy crisis by producing hydrogen fuel using sunlight and water as the only inputs [1]. Hydrogen ($H_2$), a versatile energy carrier, is believed to be one of the most promising clean energy sources that stores energy as a molecular bond. There are different kinds of devices for solar water-splitting ([2] and refs. therein). We aim to build a self-sustained PEC [3] device for sunlight-driven hydrogen production using bismuth-based metal oxides as active photoelectrode materials.

Green and renewable energy researchers have devoted tremendous efforts to developing multinary metal oxides due to inherent drawbacks in light absorption, carrier transport, and stability of binary metal oxide-based photoelectrodes. Bismuth-based metal oxides are promising candidates for photoelectrodes (for example, n-type photoanode $BiVO_4$ and p-type photocathode $CuBi_2O_4$) in solar water splitting devices since they have demonstrated excellent visible light harvesting capability and a right band edge for water splitting ([2] and refs. therein). The synthesis and physicochemical properties of n-type $BiVO_4$ photoanode materials can be found in our previous report [2]. *Kusachiite* (copper bismuth oxide), a p-type semiconductor with an ideal direct bandgap of 1.6-1.8 eV, shows great potential to be used as an active photocathode in PEC devices [1,3]. Copper bismuth oxide ($CuBi_2O_4$) is not only just eco-friendly but also its theoretical high solar-to-$H_2$ (STH) conversion efficiency, durability, and abundance on the earth, making it one of the most suitable photocathode materials. Moreover, the conduction band of $CuBi_2O_4$ (CBO) is located at a more negative potential than the reduction potential of $H^+/H_2$, allowing adequate driving force for solar $H_2$ production. Furthermore, a positive photocurrent at 1.0 V *vs* RHE makes CBO a potential photocathode material for self-sustained water-splitting systems [3,4].

There are many fabrication techniques, including wet-chemical [5], solid-state reaction [6,7], electrodeposition [8], and physical vapor methods [9] to synthesize phase pure $CuBi_2O_4$ with tunable optoelectronic properties. Ball milling, both high and low energy, has become popular for producing technologically important nanomaterials [10,11]. In this short communication, we proposed a facile top-down synthesis method with low energy sequential ball milling (LE-SBM) which can successfully produce phase pure $CuBi_2O_4$ (CBO) with slight stoichiometric variation by simply changing the milling duration. And this kind of synthesis is rarely reported in the literature. The SBM method is facile, cost-effective, and has great potential to scale up industrially compared to other existing methods in the literature [5-9]. SBM is not only suitable for producing CBO, but it also can be useful for many other complex metal oxides synthesis [2,12,13]. Additionally, this mechanochemical low-energy ball milling process has the ability to manipulate the elemental ratio i.e., Cu:Bi and O species in synthesis product to attain tunable bandgap CBO within the phase purity limit by judicious selection of processing parameters. Here, we choose LE-SBM to achieve a suitable temperature during the collision between the milling particles that favors the formation enthalpy of CBO and prevents impurity phases other than CBO [2,10]. Our top-down synthesis process yielded CBO nanopowders with the desired crystalline structure and phase purity employing LE-SBM since the continuous fragmentation and re-welding mechanism of grains happens in the process of SBM. The synthesis mechanism of phase pure CBO, with the aid of a variety of characterization techniques, was investigated systematically. Finally, preliminary photoelectrochemical (PEC) performance studies of some selective SBM-CBO samples were conducted by preparing 0.1 M slurry atop the screen printed electrode (SPE) with aqueous $Na_2SO_4$ electrolyte to assess the utility of phase pure $CuBi_2O_4$ as photocathode materials, which are discussed below.

## Experimental

### Sample preparation

In this study, the standard solid-phase reaction technique for ceramic materials [2,14] was slightly modified to synthesize CBO samples. We took analytical grade cupric oxide (CuO) (purity~99.95%) and bismuth oxide ($Bi_2O_3$) (purity~99.95%) powders as starting raw materials and pre-activated them by individually hand-milled (HM) for 1 h. The CuO and $Bi_2O_3$, taken in the molar ratio 1:1, were mixed and HM for another 1 h. Afterward, the mixture was ball-milled (BM) for 5 h at 150 rpm (low-energy ball milling) with maintaining a ball-to-powder ratio of 1:1 (Ball diameter: 10 mm, ball material: alumina). As mentioned earlier, we intentionally used low-energy ball milling to maintain sample purity and to avoid abrupt structural changes and defect formation caused by high-energy ball milling [15]. The solid mixture was stirred with a spoon every ~ 2 h to get a homogenously mixed powder. After 6 h of hand milling and ball milling (HBM), the powder was sintered at 750 °C in a furnace (model: Nabertherm P-310, Germany) over 4 h with a heating rate of 5 ºC/min and then cooled down to room temperature naturally. The resultant sample was preserved in a container after grinding (GS), and a small amount was deducted as a part of the sample collection. Then, the rest of the GS sample was further sequentially ball milled (SBM) for durations up to 25 h to attain slight stoichiometric variation of CBO within the phase purity limit and grain size reduction. Thus, samples at different ball milling hours (5, 10, 15, 20, and 25 h) were collected. We also adjusted the ball-to-powder ratio (2:1, 2.35:1, 2.60:1, 2.98:1, and ≈ 3:1) during SBM. Finally, the collected samples were preserved in airtight glass vials. The flow diagram of the whole synthesis process of CBO with sample notation can be found in Fig S1 in the supplemental information.

### Characterization techniques

*XRD analysis:* To determine the structural properties of the samples in their powder form, a Thermo Scientific ARL EQUINOX 1000 X-Ray Diffractometer (XRD) with a monochromatic copper anode source of Cu-Kα radiation (λ = 1.5406 Å) in out-of-plane geometry was used.

*Raman analysis:* A Horiva MacroRam Raman Spectrometer equipped with a 785 nm diode laser was utilized to investigate chemical structure as well as the phase purity of the CBO materials. The laser power on the sample surface was kept below 5 mW to avoid sample degradation. The equipment was calibrated with a standard Si wafer Raman peak ~520 $cm^{-1}$ prior to recording all Raman spectra at room temperature.

*FE-SEM imaging and EDX microanalysis:* The microcrystalline morphology and elemental composition of the samples were studied by an FE-SEM (JEOL 7610F) coupled with an oxford EDX analyzer (JED 2300). The grain sizes were estimated using the open-source ImageJ software.

*XPS study:* A X-ray photoelectron spectroscopy (XPS) instrument (K-ALPHA, Thermo Scientific, USA) was used to study the chemical state of the constituent elements, elemental ratios, as well as to identify the impurity phase in the synthesized CBO samples. The XPS was operated by applying a soft X-ray monochromatic source: Al Kα radiation (hν =1486.6 eV) at 15 kV and 10 mA. The residual pressure of the analysis chamber was kept below $10^{-8}$ torr. The core level C 1*s*, Cu 2*p*, Bi 4*f*, and O 1*s* spectra, after background subtraction, were subjected to curve-resolving by applying a Gaussian line shape, along with a Lorentzian broadening function. The adventitious alkyl carbon: C 1*s* signal at 284.8 eV was used as a reference to calibrate the binding energies of Cu, Bi, and O elements of the CBO samples.

*UV-Vis-NIR spectroscopy:* The optical absorption data of the samples were collected at room temperature using a UV–Vis–NIR spectrophotometer (SHIMADZU 2600 plus, Japan) coupled with an integrating sphere, using $BaSO_4$ pellet as reference for the background correction.

*Photoelectrochemical study:* The photoelectrochemical (PEC) performance of some selective samples was examined ad-hoc basis by studying the Electrochemical Impedance Spectroscopy (EIS) of 0.1M slurry samples using 0.5M $Na_2SO_4$ aqueous electrolyte using screen-printed electrode (SPE) from DropSens (DS 220AT), Spain, and data were recorded both under the dark and ~520 nm (~20 mW/cm$^2$) & ~ 365 nm (~12 mW/cm$^2$) LED illuminations. The SPE consists of a gold thin layer (dia. 4 mm, effective geometric area~0.126 cm$^2$) as the working electrode (WE), a second gold layer as the counter/auxiliary electrode (CE), and a silver thin layer as the pseudo reference electrode (RE). The cyclic voltammetry (CV) of samples was also recorded at room temperature using the same SPE and $Na_2SO_4$ under the dark. An electrochemical workstation (Autolab MetrOhm Potentiostat/Galvanostat 204) coupled with Frequency Response Analyzer (FRA32) was used for this study.

**Results and discussion**

The crystal structure, phase, and purity of the samples were analyzed from their powder X-ray diffraction (XRD) patterns shown in Fig. 1a. The XRD peaks in the Cu-Bi-O-HBM 6 h sample match neither CuO [16,17] nor $Bi_2O_3$ [18]. Hence, this presumably indicates the agglomeration state of the mixed powder state arising from mechanical alloying (MA) [19]. In contrast, the XRD pattern of CBO (750C-4 h) shows that the ternary copper bismuth oxide formation started after sintering at 750 ºC for 4 h in the air. Notice that the intensity of the dominant diffraction peak near 2θ =28° increases remarkably along with the significant shifting to the higher Bragg angles for CBO (750C-4 h) compared to those of Cu-Bi-O-HBM 6 h (see dashed vertical lines in Fig. 1a). The XRD patterns of SBM-5 h and SBM-25 h suggest that the crystallinity of CBO materials increases with increasing milling duration. Notably, most of the diffraction planes (marked by dashed lines) of the CBO (750C-4 h), SBM-5 h, and SBM-25 h samples match the XRD patterns of $CuBi_2O_4$ reported in the literature [15] and confirm the tetragonal phase with space group P4/ncc (#130) [20]. The most intense diffraction peak at about 2θ =27.99° can be assigned to the (211) lattice plane of $CuBi_2O_4$. The Full Width at Half Maximum (FWHM) of 211 Bragg peak was considered to estimate the average crystallite domain size of CBO samples by viewing a pseudo-distribution for the peak broadening and using Scherrer's formula [21]:

$$D = \frac{k\lambda}{FWHM^{sample} \cos\theta}$$

where, *D* is the crystallite domain size, $FWHM^{sample}$ is the full width at half maximum, *k* (= 0.9 for spherical grains) is Scherrer constant, λ is the wavelength of Cu-Kα radiations (1.54056 Å), and θ is the Bragg diffraction angle. The crystal domain size of Cu-Bi-O-HBM 6 h, CBO (750C-4 h), SBM-5 h, and SBM-25 h were estimated to be approximately 66.0 nm, 51.5 nm, 55.2 nm, and 50.0 nm, respectively.

In Fig. 1a, two diffraction peaks appeared at 2θ ≈ 35.480 and 37.410 may respectively match the $(11\bar{1})$ and $(\bar{2}02)$ Bragg planes of monoclinic CuO [16,17], suggesting a possible presence of a tiny amount of CuO in the CBO matrix. This was presumably because of the unreacted CuO raw powder during low-temperature sintering or produced from the decomposition of CBO during SBM processing. Similar results can also be found in refs. [21,9]. The benefits of a small amount of CuO phase segregation in CBO have also been reported [21]. The sintering temperature and time were selected intentionally much lower (< 950 $^0$C) than in the literature [7, 20,14,23-24]. We successfully reduced the sintering temperature for phase pure CBO formation, and 6 hours of HBM before sintering could be the main factor in this process.

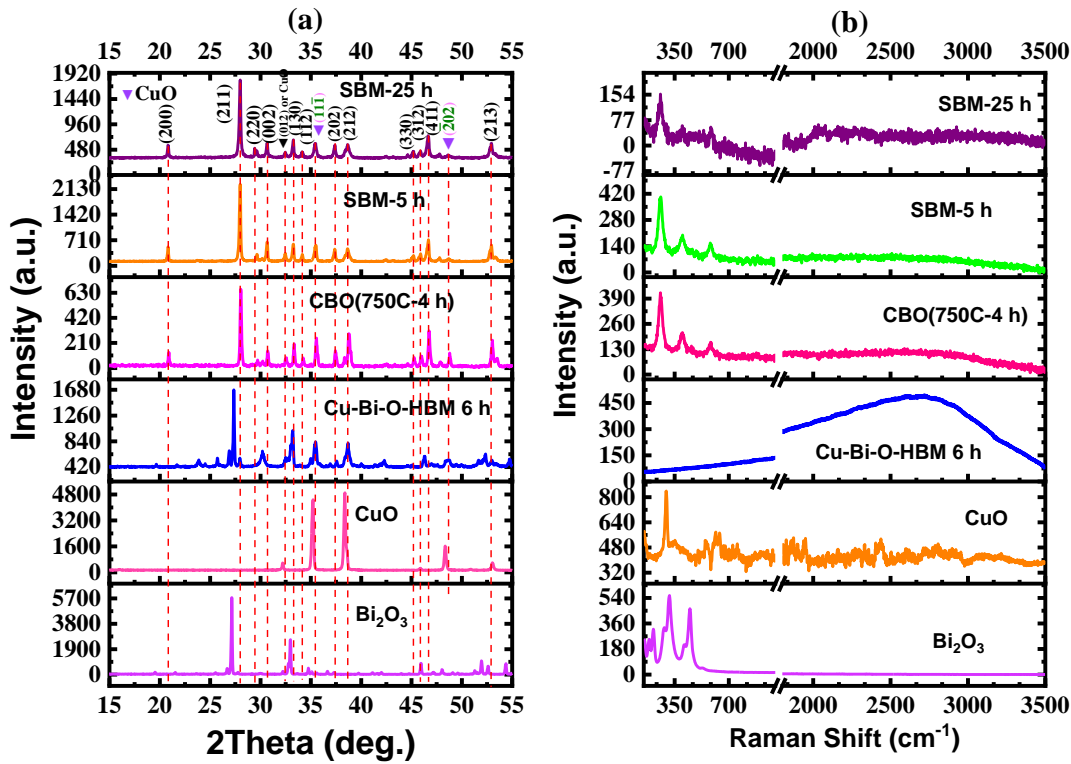

**Fig. 1** – (a) X-ray diffraction (XRD) patterns and (b) room temperature Raman spectra of the samples along with the raw materials CuO and $Bi_2O_3$.

Raman spectroscopy is a more powerful tool than XRD to assess the phase purity and grain size reduction of the samples under investigation, as the lattice vibrations are sensitive to the local atomic arrangement owing to the difference in chemical bonds and changes in atomic masses [25]. The Raman spectra of Cu-Bi-O-HBM 6 h, CBO (750C-4 h), SBM-5 h, and SBM-25 h samples are displayed in Fig. 1b. The Raman spectra of the Cu-Bi-O-HBM 6 h sample is featureless with a broad amorphous-like hump and also match neither with CuO [16,26] nor with $Bi_2O_3$ [18] corroborating the XRD results discussed above. It may imply that during the course of hand-ball milling (HBM) of the raw materials for 6 hours, the particles of individual raw materials may come in contact with each other at the time of ball-to-ball and ball-to-vial wall collision; nucleation of the particles then takes place by solid-state diffusion and forms an amorphous alloy [10]. In contrast, three Raman bands centering at about 259.6 $cm^{-1}$, 399.6 $cm^{-1}$, and 583.3 $cm^{-1}$ are observed for CBO (750C-4 h), SBM-5 h, and SBM-25 h samples without any detectable vibrational peaks of CuO and $Bi_2O_3$. These Raman bands agree with the vibrational structure of tetragonal $CuBi_2O_4$ reported previously [27,28]. This observation confirms that phase pure CBO could be grown from raw CuO and $Bi_2O_3$ powders using a facile SBM technique. Furthermore, the gradual peak broadening (for example, the 259.6 $cm^{-1}$ peak) of the samples indicates grain size reduction is due to SBM with increasing milling durations. Both stochiometric changes and grain size reduction due to SBM may affect the optical bandgap of the synthesized CBO, which will be discussed in the following sections.

SEM micrographs and elemental compositions of SBM-5 h and SBM-25 h samples are shown in Fig. 2a and 2b, revealing average grain sizes of ~145 nm and ~102 nm, respectively. The grain size of a sample is sensitive to the milling conditions, such as time and rotation rate (or energy type). The grain size reduction with the SBM duration is clearly evident from the SEM micrographs. In Fig. 2, EDX spectra of the SBM-5 h and SBM-25 h samples exhibit the signatures of the constituent elements (Cu, Bi, and O) of $CuBi_2O_4$. The detected chemical species and their percentages (atomic and weight) are enlisted along with the corresponding EDX spectra (see the inset of the bottom panel of Fig. 2a and 2b). The stoichiometric changes with SBM durations are evident in both samples, corroborating the results reported by others [10, 29]. However, it can be concluded that SBM-25 h samples are comparatively phase pure than those of SBM-5 h samples from EDX and XRD data. The Cu/Bi ratio, collected from EDX data, in SBM-5 h and SBM-25 h are ~1.8 and ~2, respectively. Since it is well known that EDX is a semi-quantitative elemental analysis, therefore, one of the samples (SBM-25 h) was characterized by a more sensitive XPS technique to confirm further its phase purity and chemical composition.

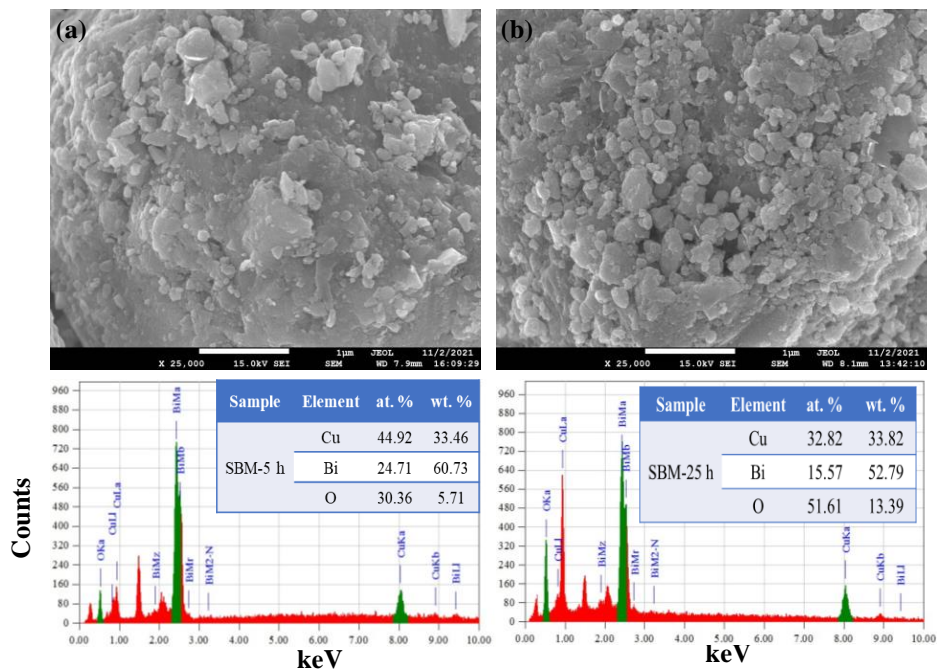

**Fig. 2** – SEM micrographs and EDX spectra of the (a) SBM-5 h and (b) SBM-25 h samples.

The survey spectrum of SBM-25 h and raw $Bi_2O_3$ powder sample is presented in Fig. 3a. The survey spectra were recorded from 0-1050 eV. Only C 1$s$ was found as an adventitious carbon contaminant at ~284.8 eV binding energy for both SBM-25 h and $Bi_2O_3$ samples. The undefined peaks may exhibit electron energy loss features due to plasmon excitation [21]. Splitting observed in the peaks due to 4$d$ and 4$f$ orbitals of Bi while splitting in Bi 4$d$ is more prominent than in Bi 4$f$. Not only just Bi 5$d$, Bi 4$f$, Bi 4$d$, Bi 4$p$, Cu 2$p$, C 1$s$, and O 1$s$ photoelectron lines but also the CuLM1 and CuLM2 Auger lines are found in the SBM-25 h sample. The high-resolution XPS spectra of Bi 4$f$, Cu 2$p$, and O 1$s$ are illustrated in Fig. 3b-3d, and their corresponding BE peaks are summarized in Table 1. These peaks attribute to the oxidation state and spin-orbital state of the core atoms. Two strong peaks of Bi 4$f$ were observed at 158.81 eV and 164.12 eV (for SBM-25 h), corresponding to doublet spin-orbital photoelectron emission from Bi $4f_{7/2}$ and Bi $4f_{5/2}$ states indicating the presence of $Bi^{3+}$ oxidation state in the materials. Higher shifting in BE for SBM-25 h than $Bi_2O_3$ indicates phase pure CBO formation in the SBM-25 h sample with oxidation state $Bi^{3+}$ [30]. The high-resolution XPS spectra of Cu 2$p$ in SBM-25 h (Fig. 3c) manifests two main peaks centered at 932.64 and 952.43 eV, with a horizontal separation of 19.79 eV, corresponding to the photoelectron emission from Cu $2p_{3/2}$ and Cu $2p_{1/2}$ microstates of pure $CuBi_2O_4$, respectively [31]. Besides, few satellite peaks are found in Cu 2$p$ spectra of SBM-25 h. The Cu 2$p$ peak positions and the presence of satellite peaks confirm that the copper species exists in the +2-oxidation state ($Cu^{2+}$) [32]. These satellite peaks (marked by fade vertical lines) in SBM-25 h samples significantly differ from those of the CuO and $Cu_2O$ phases, suggesting their successful formation in phase pure $CuBi_2O_4$. Furthermore, the Bi/Cu ratio estimated from the XPS survey of SBM-25 h is ~1.7, which is close to the 2 for the ideal $CuBi_2O_4$.

The O 1$s$ XPS signals shown in Fig. 3d are deconvoluted into three peaks separately for SBM-25 h and $Bi_2O_3$ samples. The predominant peak of O 1$s$, 529.52 eV and 529.62 eV for $Bi_2O_3$ and SBM-5 h, respectively, can be ascribed to the crystal lattice oxygen ($O_L$) [32]. The other O 1s peaks with higher binding energies belong to the oxygen vacancy ($O_V$) and oxygen absorption ($O_{ab}$), respectively. These peaks could arise due to surface defects or chemisorbed oxygen species [33]. The peak energy shift towards higher BE for SBM-25 h is also prominent. In summary, XPS analyses confirmed that slight stoichiometric variation within the phase purity limit of $CuBi_2O_4$ could be achieved by facile low-energy SBM technique, which is desirable for tunable bandgap photocathode for PEC devices [3].

**Table 1.** Measured binding energies (in eV) of Bi 4f, Cu 2p, and O 1s photoelectron lines in SBM-25 h, $Bi_2O_3$, CuO, and $Cu_2O$.

| Materials | Bi 4f | Cu 2p |
|---|---|---|
| SBM-25 h | 158.81, 164.12 | 932.64, 952.43 |
|  | 158.4, 163.8 [31] | 934.2, 954.4 [31] |
|  | 158.6, 163.9 [32] | 934.1, 953.8 [32] |
| $Bi_2O_3$ | 158.70, 164.00 | - |
| CuO | - | 933.13, 953.09, 961.94 |
| $Cu_2O$ | - | 932.68, 952.60 |

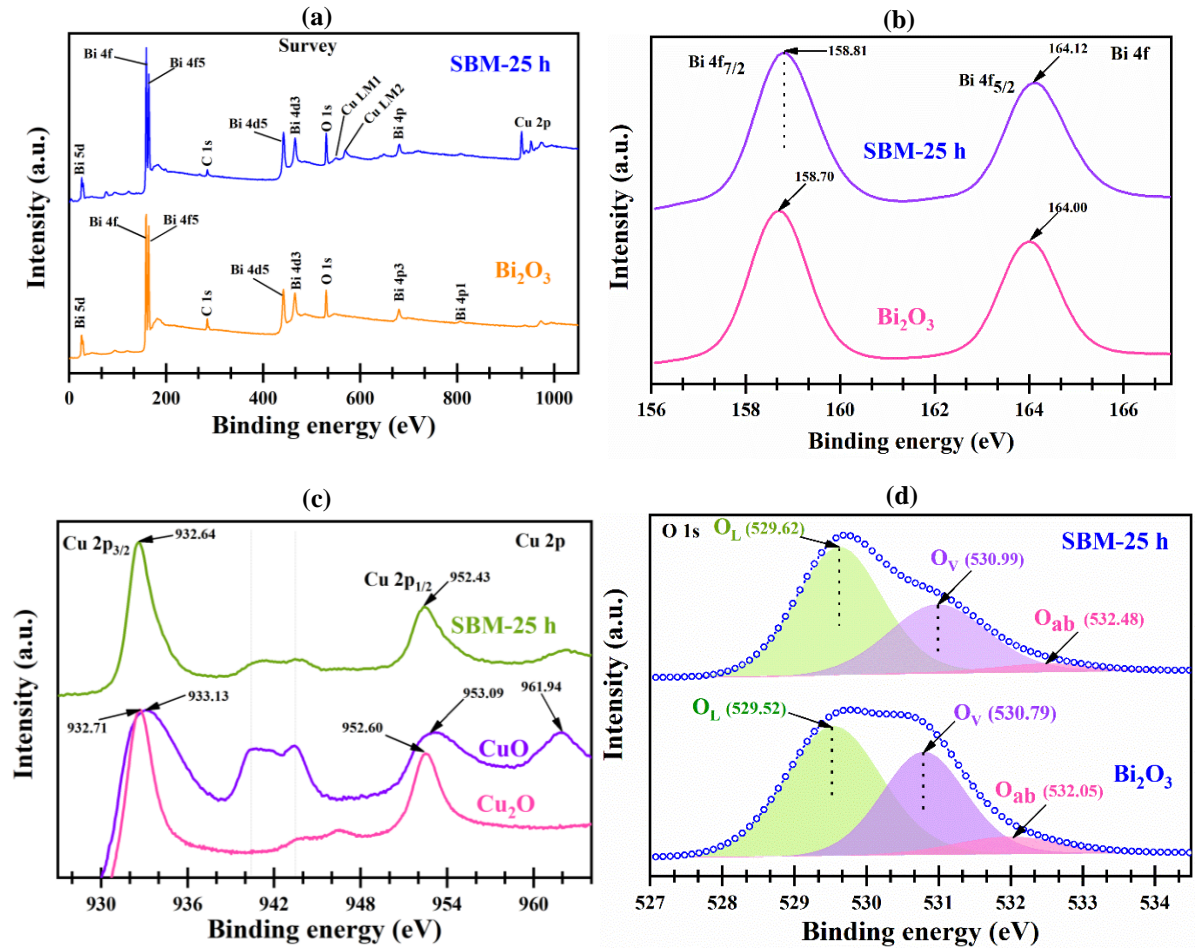

**Fig. 3** – (a) XPS spectra of $CuBi_2O_4$ and $Bi_2O_3$ showing regions for the core level of (b) Bi 4f, (c) Cu 2p, and (d) O 1s.
Page 7 of 12

To get insight into the light-harvesting capability of the CBO, the diffuse reflection data of SBM powder samples were converted to absorption spectra by Kubelka-Munk (KKM) function [2] and displayed in Fig. 4a. The absorption edge wavelength (energy) of raw CuO and $Bi_2O_3$ powders is around 900 nm (~1.38 eV) and 437 nm (~2.84 eV), respectively. Notably, the primary absorption edges of synthesized CBO samples are located within the 650-850 nm wavelength range. The absorption edge variation (different from CuO and $Bi_2O_3$) is presumably due to the combined effect of grain size and elemental concentration variation in CBO. Many parameters, such as grain shape and size, the presence of different vacancies and interstitials, and defects, controls optical absorption [2]. For instance, the red shift in the absorption edge usually comes from the grain size increment [34]. The optical properties of nanostructured materials strongly depend on the prepared samples' microstructure. It has also been reported in the literature that ball-milling time influences the stoichiometric ratio of the material [29], which could influence the optical bandgap of materials under study and, thereby, its photoelectrochemical performance, as discussed below.

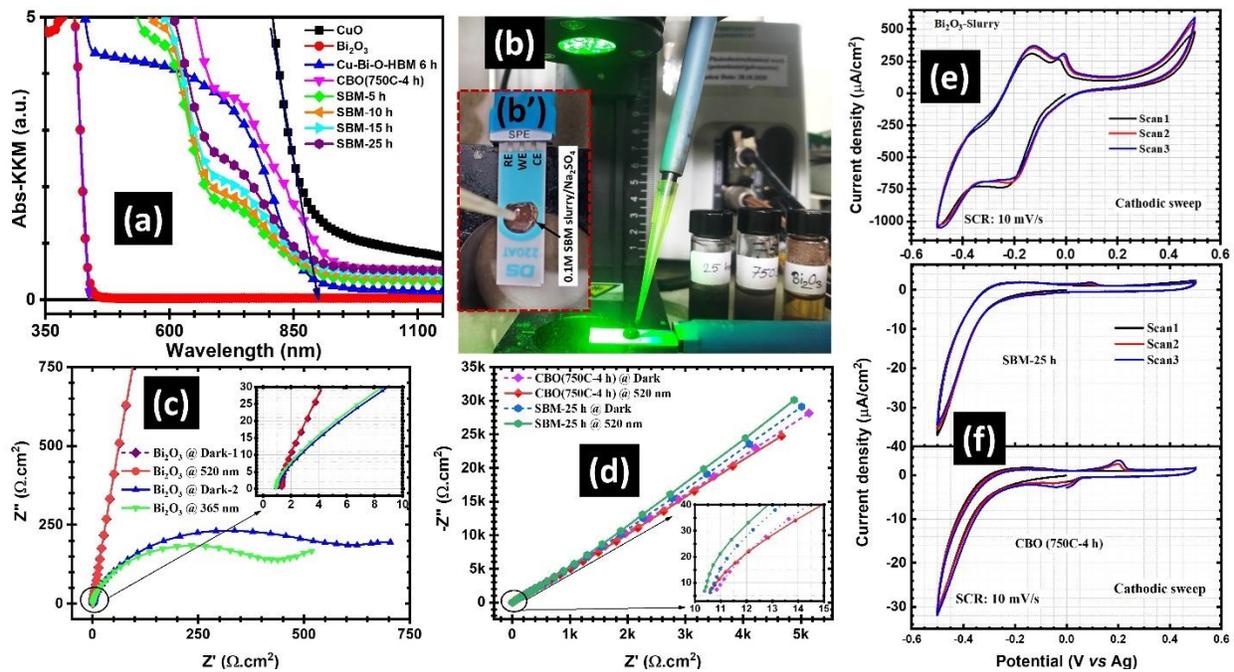

**Fig. 4** – (a) UV-Vis diffuse absorption spectra of raw CuO, raw $Bi_2O_3$, and synthesized CBO powders with different milling conditions, (b) Experimental setup for measuring photoinduced-EIS and Cyclic voltammograms of 0.1M slurry samples using 0.5 M $Na_2SO_4$ electrolyte with the screen printed electrode (SPE), the inset (b') shows a close view of the slurry droplet (~50 µL) atop the gold (Au) SPE (DS-220AT), (c) Nyquist plots of 0.1 M $Bi_2O_3$ under the dark and 520 nm & 365 nm LED illumination; (d) EIS of CBO(750C-4 h) and SBM-25 h samples under the dark and 520 nm LED illumination; the insets of (c) and (d) shows the enlarged part of the high-frequency region of the respective EIS; Cyclic voltammograms of (e) $Bi_2O_3$, and (f) CBO(750C-4 h) & SBM-25 h recorded at room temperature immediately after the photoinduced-EIS measurements.

A preliminary photoelectrochemical (PEC) performance study of some selective samples was performed by preparing 0.1 M slurry samples using 0.5 M aqueous $Na_2SO_4$ electrolyte placed on the Au-SPE by the EIS technique under the dark and LED illuminations (see Fig. 4b and 4b'). The room temperature EIS measurements were carried out in the frequency range of 100 mHz – 500 kHz with a small *AC* perturbation amplitude of ~10 mV, and the Nyquist plots were constructed from the Z' (real) and Z" (imaginary) part of impedances (EIS) data. Figure 4c shows the Nyquist plots of 0.1 M $Bi_2O_3$ under the dark and 365 nm (~2.84 eV) & 520 nm (~2.38 eV) LED illuminations. The Nyquist plots recorded under the dark consist of a broad semi-circle at the high-frequency region followed by a straight line in the low-frequency region (dark-2, blue curve). The straight lines in the low-frequency region of Nyquist plots are ascribed to the ion(s) diffusion of electrolyte into/from the sample-electrode, while semi-circles at the high-frequency region are attributed to the charge-transfer resistance at the sample-electrode|electrolyte interface ([35] and refs. therein). Notice that under the 365 nm (~2.84 eV) LED illumination, the impedance of Nyquist plots is evidently decreased, and the starting point of the semi-circles (green and blue curves) clearly shifted to the lower impedance region (see inset of Fig. 4c) due to the illumination photon energy being higher than the bandgap energy of $Bi_2O_3$ ($E_g$~2.84 eV). At the mid- to low-frequency regions, the dark curves are seen to be raised compared to their illuminated counterparts, presumably due to the competing process between the ion-diffusion and (photoinduced)charge-transfer processes. Notice also that under the dark and 520 nm (~2.38 eV) LED illumination, the Nyquist plots and their starting points remain the same for $Bi_2O_3$ (see also Fig. S2 in the supplementary information). This study implies that the selection of illumination photon energy above the bandgap of slurry samples may be utilized to assess the PEC performance quickly using SPE. Similarly, to compare the PEC quality of synthesized $CuBi_2O_4$ photocathode materials, the EIS spectra of CBO(750C-4 h) and SBM-25 h samples are recorded under the dark and 520 nm LED illumination and shown in Fig. 4d. Notice that under 520 nm LED illumination, starting points of semi-circles (solid curves) are shifted to the low Z' impedance values compared to their dark counterparts (dashed curves) (displayed in the inset of Fig. 4d). Evidently, the illuminated-Nyquist plots for the SBM-25 h samples consistently remains above the dark-Nyquist plots throughout the whole frequency region, but for CBO(750C-4 h) samples, the dark Nyquist plots go above the illuminated one after ~12 $\Omega.cm^2$ (see Fig. 4d inset). This can be attributed to the enhanced photogenerated charge transfer and electro-catalytic ability of SBM-25 h samples compared to CBO(750C-4 h) samples due to better phase purity and crystalline quality of the former compared to the latter. We also recorded the cyclic voltammogram (CV) of the same 0.1 M slurry samples (i.e., $Bi_2O_3$, CBO(750C-4 h), and SBM-25 h) immediately after illuminated EIS measurements to assess their possible photodegradation or materials quality. Firstly, cathodic sweep (i.e., negative potential scanning direction from the starting point) was employed in CV recording to polarize the slurry to the Au-working electrode (SPE-Au) with a slow scanning rate of 10 mV/s shown in Fig. 4e and 4f. The observed oxidation (OP) and reduction (RP) peaks in three consecutive CV scans for $Bi_2O_3$ (see Fig. 4e) are found to be similar in nature as reported in the literature [36,37], but the potentials are different as we used Au-SPE for this study. CVs of 0.1 M $Bi_2O_3$ supernatant (~50 µL) droplet atop the Au-SPE under the dark were also recorded at room temperature for comparison purposes (see Fig. S3, Fig. S4, Table S1, and Table S2 in the supplementary information file). Intriguingly, CVs of CBO (750C-4 h) and SBM-25 h samples exhibit different OP and RP compared to those of $Bi_2O_3$ (cf. Fig. 4e and 4f). Clearly, voltammograms of SBM-25 h samples are comparatively featureless amongst the samples recorded immediately after EIS measurements. Therefore, the photoelectrochemical performance of SBM-25 h samples in aqueous $Na_2SO_4$ electrolyte is superior to other samples studied due to their good stability, phase purity, and good (photo)electrocatalytic properties. Further experiments with thin film-based photocathode of these samples are in progress to evaluate their $H_2$ generation efficiency and will be reported elsewhere.

## Conclusion

In summary, a facile top-down synthesis with low-energy sequential ball milling (LE-SBM) has been successfully developed for the production of phase pure tetragonal $CuBi_2O_4$ nanopowder. Our synthesis process employs a lower sintering temperature and time (i.e., 750º C and 4 h), compared to the process reported in the literature. The phase pure $CuBi_2O_4$ was confirmed in the 25 h sequentially ball-milled (SBM-25 h) samples from the combined analyses of XRD, Raman, and XPS. FE-SEM micrographs confirmed the grain size reduction with SBM durations. The formation of $CuBi_2O_4$ (CBO) has also been verified from the elemental composition conducted by FE-SEM/EDX analyses. The EDX and XPS data reveal that the stoichiometric ratio (Bi:Cu) changes with ball milling time. In addition, the optical absorption studies revealed that the CBO with a tunable optical bandgap (~1.70 –1.85 eV) was achieved by LE-SBM. Finally, preliminary photoelectrochemical performance studies of synthesized samples revealed that 0.1 M SBM-25 h slurry atop the gold screen-printed electrode with aqueous $Na_2SO_4$ electrolyte was superior to other samples due to their good stability, phase purity, and good (photo)electrocatalytic properties.

## CRediT authorship contribution statement

**S.F.U.F.**: Conceptualization, Methodology, Investigation, Data curation, Formal analyses, Supervision, Funding acquisition, Writing- original draft, Writing – review & editing. **M.I.N.**: Data curation, Formal analyses, Writing – original draft, Writing - review & editing. **N.I.T.**: Investigation, Data curation, Formal analyses; **M.S.Q.**: XPS investigation, visualization; **S.A.J.**: Visualization, resources; **S.I.**: Visualization, resources; **M.A.A.S.**: Visualization, resources, validation.

## Declaration of competing interest

The authors declare that they have no known competing financial interests or personal relationships that could have appeared to influence the work reported in this paper.

## Acknowledgements

All the authors are thankful to the Industrial Physics Division (IPD), BCSIR Dhaka Laboratories, Dhaka 1205, and extend their appreciation for financial and Instrumental support under the scope of R&D project# TCS-FY2017-2022, Special Allocation Grant# 404-ES-FY2021-22 (Ref.#39.00.0000.009.14.019.21-745 Dated 15/12/2021), Ministry of Science and Technology, Government of Bangladesh, the World Academy of Sciences (TWAS) Grant#20-143 RG/PHYS/AS_I, and Royal Society of Chemistry, UK Grant# R20-3167 for IPD. A special thanks to BTRI of BCSIR for helping with FE-SEM/EDX of CBO samples.